\documentclass{aa}
\usepackage{graphicx}
\begin{document}

\title{The limb darkening  of $\alpha$\,Cen~B}
\subtitle{Matching 3D hydrodynamical models with interferometric
measurements}
\author {L. Bigot \inst{1}
        \and
        P. Kervella \inst{2}
        \and
        F. Th\'evenin\inst{1}
        \and
        D. S\'egransan\inst{3}}
\offprints{L. Bigot}
\institute{D\'epartement Cassiop\'ee, UMR 6202, Observatoire de la
C\^ote d'Azur, BP 4229, 06304 Nice Cedex 4, France \and LESIA, UMR
8109, Observatoire de Paris-Meudon, 5 place Jules Janssen, 92195
Meudon Cedex, France \and Observatoire de Gen\`eve, CH-1290
Sauverny, Switzerland}
\titlerunning{The limb darkening of $\alpha$\,Cen\,B}
\authorrunning{Bigot et al.}
\mail{lbigot@obs-nice.fr}
\date{Received date / Accepted date}

\abstract{ For the nearby dwarf star $\alpha$\,Cen~B  (K1\,V), we
present limb darkening predictions from a 3D hydrodynamical
radiative transfer model of its atmosphere.
We first compare the results of this model to a standard Kurucz's
atmosphere. Then we use both predictions to fit the new
interferometric visibility measurements of $\alpha$\,Cen~B obtained
with the VINCI instrument of the VLT Interferometer.
Part of these new visibility measurements were obtained in the
second lobe of the visibility function, that is sensitive to stellar
limb darkening.
 The best agreement is found for the 3D atmosphere limb darkening
model and a limb darkened angular diameter of $\theta_{\rm 3D} =
6.000\pm 0.021$\,mas, corresponding to a linear radius of $0.863 \pm
0.003$\,R$_\odot$ (assuming $\pi = 747.1 \pm 1.2$\,mas).
Our new linear radius is in good agreement with the asteroseismic
value predicted by Th\'evenin et al.~(\cite{the02}). In view of
future observations of this star with the VLTI/AMBER instrument, we
also present limb darkening predictions in the $J$, $H$ and $K$
bands. \keywords{Stars: individual: $\alpha$\,Cen, Techniques:
interferometric, Stars: binaries: visual, Stars: fundamental
parameters, Stars: atmospheres} } \maketitle

\section{Introduction}\label{sec:int}

The limb darkening (hereafter LD) is a well known effect in stellar
physics. Its manifestation is  a non-uniform brightness  of the disk
whose  edges appear fainter than the center. This effect occurs
because of the decrease of the source function outwards in the
atmosphere. The disk center then shows deeper and warmer layers
whereas the edges show higher and cooler material. This means that
the analysis of the intensity $I_{\lambda} (\mu)$ at different
latitudinal angles $\mu=\cos\theta$ provides information on the
temperature variation with depth in the external layers of the star.
This is therefore an excellent constraint to test atmospheric
models, validate or invalidate assumptions used to derive these
models (like NLTE/LTE) and provide hints for improving the input
physics (equation-of-state and/or opacities in particular). The
center-to-limb variation of the Sun is known for many years and has
been measured for numerous $\mu$ and $\lambda$ (e.g. Pierce \&
Slaughter 1977, Neckel \& Labs 1994, Hestroffer \&
Magnan~\cite{hestroffer98}) leading to a plethora of theoretical
works which have improved our knowledge of the external layers of
the Sun.

Traditionally, the analysis of solar and stellar LD is made by
adopting an approximated law for $I_\lambda (\mu)$, generally a
polynomial expansion in $\mu$, either linear or non-linear (see e.g.
Claret~\cite{claret00} for recent developments), whose coefficients
are determined from 1D atmospheric models, like ATLAS
(Kurucz~\cite{kurucz92}) or Phoenix
(Hauschildt~\cite{hauschildt99}). However, in spite of the detailed
physics included in these codes, their 1D nature is a limitation for
deriving realistic emergent intensities. Indeed, these codes contain
free parameters like the well-know mixing length parameter that are
injected artificially in order to reproduce the properties of the
turbulent convection at the stellar surface. As a consequence, the
comparison between these 1D models and observations depend on the
input parameters that thereby creates an important source of
uncertainties. Moreover, the convection is by nature a 3D process.
Its manifestation is  the  presence of bright granules and dark
intergranular lanes. Reducing it to a 1D process, i.e. ignoring
horizontal flows and temperature inhomogeneities,  changes the
pressure scale height, the location of the surface and therefore
also the emergent intensity (see e.g. Allende Prieto, Asplund \&
Fabiani Bendicho \cite{allende04} and Asplund et al.
\cite{asplund00a} for a comparison between multi-dimensional
simulations).

The precise measurements of the center-to-limb variation achieved
nowadays require realistic stellar atmospheric models that take into
account all the complexity of the stellar surface, and motivates the
use of the new generation of 3D radiative hydrodynamical (hereafter
RHD) simulations.

In this paper we propose a study of $\alpha$\,Cen~B
(\object{HD128621}), a nearby  K1V dwarf star. It is part of a
visual triple star system whose brightest component, $\alpha$\,Cen~A
(\object{HD128620}), is a G2V dwarf. The motivation for the
selection of this star in the present work lies in both theoretical
and observational considerations. From the interferometric point of
view, the proximity ($1.3$\,pc) of the star is a rare opportunity to
allow interferometric measurements since most of the nearby dwarfs
have  too small angular diameters to be measured. Our interest for
this star has grown recently since our new measurements provide data
points in the second lobe of the visibility function which is
sensitive to the LD of the star. From a theoretical point of view,
this star is important for various reasons. In particular, the
recent detection of solar-like oscillations in $\alpha$\,Cen~A and B
(Bouchy \& Carrier~\cite{bouchy01}, ~\cite{bouchy02}, Carrier \&
Bourban ~\cite{carrier03}) have led several authors (e.g. Morel et
al.~\cite{morel00}, Th\'evenin et al.~\cite{the02}, Thoul et
al.~\cite{thoul04}, Eggenberger et al.~\cite{egge04}) to build
evolution models of these two stars that are strongly constrained by
the measured frequency spacings. The result is a better, but still
debated determination of the fundamental parameters of the system.

In Sect.~\ref{obs}, we report the new interferometric measurements
of $\alpha$\,Cen~B obtained since 2003 using the VINCI instrument.
Sect.~\ref{simu} describes our 3D simulations to derive
self-consistent stellar limb darkening of $\alpha$\,Cen~B. They are
subsequently used to compute visibility curves in the near-infrared
(Sect.\ref{discussion}) in order to interpret our measurements in
terms of stellar angular diameter and to discuss the agreement
between 3D limb darkening model and our second lobe visibility
measurements. We also use our simulations to predict future
observations ($J$, $H$, and $K$ bands) that will be made with the
next generation of instruments of the VLTI, such as the new AMBER
instrument (Petrov et al.~\cite{petrov00}, Robbe-Dubois et
al.~\cite{robbe04}).

\section{New interferometric observations\label{obs}}

\begin{table*}
\caption{Calibration sequence of $\alpha$\,Cen\,B on the B3-M0
baseline (140\,m ground length). The expected visibilities $V^2_{\rm
theo}$ given in this table include the bandwidth smearing effect.
The interferometric efficiency given in italic characters
corresponds to the value assumed for the calibration of these
particular $\alpha$\,Cen\,B observations (see
Tables~\ref{table_alfcenB_1} and \ref{table_alfcenB_2}). The data of
HR\,4831 marked with ($^{*}$) were taken 2\,h before
$\alpha$\,Cen\,B. They are listed to show the stability of the IE,
but were not used for the IE estimation.} \label{Cal_alfcen}
\begin{tabular}{ccccccll}
\hline
JD & Scans & B\,(m) & Azim. & $\mu^{2} \pm$ stat. (\%) & $V^2_{\rm theo} \pm$ syst. (\%) & IE $\pm$ stat. $\pm$ syst. (\%) & Target\\
\hline
2452770.5474 & 365 & 139.309 & 49.24 & $29.37 \pm 0.39$ & $49.93 \pm 0.80$ & $58.83 \pm 0.79 \pm 0.94^{*}$ & HR\,4831$^{*}$\\
2452770.5523 & 316 & 139.131 & 50.51 & $29.38 \pm 0.51$ & $50.06 \pm 0.80$ & $58.70 \pm 1.02 \pm 0.93^{*}$ & HR\,4831$^{*}$\\
2452770.5572 & 296 & 138.913 & 51.77 & $30.05 \pm 0.67$ & $50.19 \pm 0.80$ & $59.88 \pm 1.34 \pm 0.95^{*}$ & HR\,4831$^{*}$\\
\hline
2452770.6368 & 405 & 128.934 & 71.28 & $32.05 \pm 0.35$ & $55.48 \pm 0.75$ & $57.78 \pm 0.62 \pm 0.77$ & HR\,4831\\
2452770.6419 & 408 & 127.841 & 72.49 & $31.60 \pm 0.36$ & $56.01 \pm 0.74$ & $56.42 \pm 0.64 \pm 0.74$ & HR\,4831\\
2452770.6469 & 392 & 126.698 & 73.69 & $32.72 \pm 0.36$ & $56.54 \pm 0.73$ & $57.87 \pm 0.64 \pm 0.74$ & HR\,4831\\
\hline
2452770.6605 & 299 & 133.838 & 59.85 & $0.791 \pm 0.039$ & & $\it 57.36 \pm 0.82 \pm 0.74$ & $\alpha$\,Cen\,B \\
2452770.6656 & 235 & 133.277 & 61.33 & $0.777 \pm 0.082$ & & $\it 57.36 \pm 0.82 \pm 0.74$ & $\alpha$\,Cen\,B \\
\hline
\end{tabular}
\end{table*}

\begin{figure}
\includegraphics[bb=0 0 360 288, width=8.5cm]{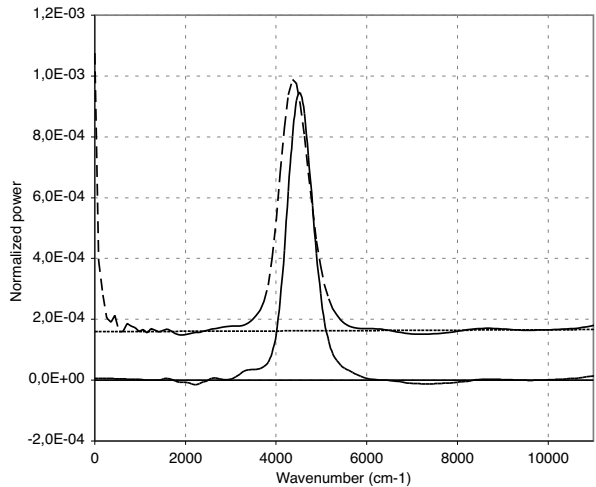}
\caption{  Average wavelets power spectral density (WPSD) of 299
interferograms of $\alpha$\,Cen\,B obtained on JD 2452770.6605 (11
May 2003). In spite of the very low visibility ($V^2 = 1.38$\,\%),
the subtraction of the background noise (dotted line) from the
processed fringes power peak (dashed line) leaves no residual bias
on the final WPSD (solid line). The power integration is done
between wave numbers 1970 and 7950~cm$^{-1}$. } \label{wl_psd_peak}
\end{figure}

A total of 37 new interferometric measurements of $\alpha$\,Cen\,B
have been obtained in 2003 on two baselines, D1-B3 (24\,m in ground
length) and B3-M0 (140\,m), using the VINCI instrument (Kervella et
al.~\cite{kervella00}; Kervella et al.~\cite{kervella03a}). The
points obtained on the longer baseline are located in the second
lobe of the visibility function of $\alpha$\,Cen\,B, whose shape
depends in the limb darkening.
We have obtained 1\,000 interferograms on the B3-M0 baseline, in two
series. Out of these, 534 were processed by the VINCI pipeline. The
B3-M0 baseline observations are made difficult by the very low $V^2$
of the interferometric fringes, less than 2\%. However,
Fig.~\ref{wl_psd_peak} shows as an example the power spectral
density of these very low visibility fringes where no bias is
present. On the D1-B3 baseline, we have recorded 17500
interferograms in 35 series (15141 processed). These new
measurements were added to the $V^2$ values obtained on the E0-G1
baseline and already published in Kervella et
al.~(\cite{kervella03b}, hereafter Paper I). The resulting squared
visibilities are listed in Tables~\ref{table_alfcenB_1} and
\ref{table_alfcenB_2}.

We have used several stars from the Cohen et al.~(\cite{cohen99})
catalog as calibrators to estimate the point source response of the
interferometer. They were observed immediately before or after
$\alpha$\,Cen\,B. On the D1-B3 baseline, we have used HD\,119193
($\theta_{\rm UD} = 2.03 \pm 0.022$\,mas), 58\,Hya ($\theta_{\rm UD}
= 3.13 \pm 0.030$\,mas) and HD\,112213 ($\theta_{\rm UD} = 3.14 \pm
0.025$\,mas). Approximately one third of the measurements was
obtained with each of these calibrators. On the B3-M0 baseline, we
have relied on HR\,4831 ($\theta_{\rm UD} = 1.66 \pm 0.018$\,mas),
whose small size results in a low systematic uncertainty on the
calibrated $V^2$ values. The angular diameter estimates from Cohen
et al.~(\cite{cohen99}) have been verified by Bord\'e et
al.~(\cite{borde02}) to be reliable within their stated error bars.
The squared visibilities were derived using the processing methods
described in Kervella et al.~(\cite{kervella04}). As an example, the
calibration sequence used for the longest baseline B3-M0 is
presented in Table~\ref{Cal_alfcen}.

\begin{table}\caption{$\alpha$\,Cen B squared visibilities.}
\label{table_alfcenB_1}

\begin{tabular}{lccc}

\hline

JD & B (m) & Azim. & $V^{2} \pm$ stat. $\pm$ syst. (\%)\\
\hline
D1-B3\\
2452720.9141 & 20.891 & 108.36 & 81.24 $\pm$ 3.03 $\pm$ 0.09\\
2452720.9081 & 21.095 & 106.30 & 83.40 $\pm$ 2.80 $\pm$ 0.09\\
2452725.8927 & 21.152 & 105.72 & 80.48 $\pm$ 1.83 $\pm$ 0.04\\
2452720.9029 & 21.270 & 104.52 & 83.89 $\pm$ 2.66 $\pm$ 0.09\\
2452725.8878 & 21.315 & 104.05 & 79.99 $\pm$ 1.81 $\pm$ 0.04\\
2452725.8828 & 21.479 & 102.36 & 80.95 $\pm$ 1.84 $\pm$ 0.04\\
2452720.8627 & 22.462 & 91.40 & 83.75 $\pm$ 3.84 $\pm$ 0.10\\
2452725.8408 & 22.669 & 88.80 & 79.26 $\pm$ 2.54 $\pm$ 0.06\\
2452725.8358 & 22.786 & 87.24 & 78.79 $\pm$ 2.52 $\pm$ 0.06\\
2452720.8489 & 22.799 & 87.08 & 82.95 $\pm$ 3.88 $\pm$ 0.10\\
2452725.8306 & 22.903 & 85.63 & 79.32 $\pm$ 2.54 $\pm$ 0.06\\
2452720.8434 & 22.921 & 85.36 & 82.06 $\pm$ 3.76 $\pm$ 0.10\\
2452726.8032 & 23.375 & 77.99 & 77.19 $\pm$ 0.64 $\pm$ 0.07\\
2452703.8642 & 23.405 & 77.43 & 80.79 $\pm$ 2.25 $\pm$ 0.05\\
2452726.7983 & 23.452 & 76.49 & 78.82 $\pm$ 0.60 $\pm$ 0.07\\
2452703.8599 & 23.470 & 76.13 & 80.05 $\pm$ 2.23 $\pm$ 0.05\\
2452726.7933 & 23.525 & 74.96 & 77.69 $\pm$ 0.59 $\pm$ 0.07\\
2452703.8555 & 23.534 & 74.77 & 81.05 $\pm$ 2.23 $\pm$ 0.05\\
2452723.7937 & 23.627 & 72.58 & 77.60 $\pm$ 0.93 $\pm$ 0.10\\
2452723.7885 & 23.688 & 71.00 & 77.11 $\pm$ 0.86 $\pm$ 0.10\\
2452723.7835 & 23.741 & 69.46 & 78.76 $\pm$ 0.81 $\pm$ 0.10\\
2452723.7521 & 23.953 & 59.81 & 78.42 $\pm$ 0.77 $\pm$ 0.10\\
2452703.8019 & 23.970 & 58.32 & 79.61 $\pm$ 2.50 $\pm$ 0.05\\
2452723.7469 & 23.970 & 58.20 & 78.20 $\pm$ 0.83 $\pm$ 0.10\\
2452704.7984 & 23.971 & 58.09 & 81.26 $\pm$ 0.98 $\pm$ 0.05\\
2452703.7979 & 23.980 & 57.06 & 80.38 $\pm$ 2.46 $\pm$ 0.05\\
2452704.7940 & 23.982 & 56.70 & 81.38 $\pm$ 0.97 $\pm$ 0.05\\
2452723.7419 & 23.982 & 56.63 & 77.93 $\pm$ 0.55 $\pm$ 0.10\\
2452709.7555 & 23.985 & 48.87 & 82.93 $\pm$ 3.59 $\pm$ 0.10\\
2452704.7896 & 23.989 & 55.34 & 80.51 $\pm$ 1.01 $\pm$ 0.05\\
2452716.7402 & 23.990 & 50.10 & 77.55 $\pm$ 3.22 $\pm$ 0.08\\
2452709.7596 & 23.991 & 50.19 & 81.43 $\pm$ 3.55 $\pm$ 0.10\\
2452726.7251 & 23.994 & 53.95 & 76.95 $\pm$ 0.90 $\pm$ 0.08\\
2452716.7448 & 23.994 & 51.55 & 76.31 $\pm$ 3.22 $\pm$ 0.08\\
2452709.7640 & 23.994 & 51.57 & 77.31 $\pm$ 3.73 $\pm$ 0.09\\
\hline
\end{tabular}
\end{table}

\begin{table}\caption{$\alpha$\,Cen B squared visibilities (continued).}
\label{table_alfcenB_2}
\begin{tabular}{lccc}
\hline
JD & B (m) & Azim. & $V^{2} \pm$ stat. $\pm$ syst. (\%)\\
\hline
E0-G1$^{\mathrm{*}}$\\
2452462.5836 & 60.441 & 157.57 & 17.02 $\pm$ 0.36 $\pm$ 0.26\\
2452462.5870 & 60.544 & 158.40 & 17.01 $\pm$ 0.23 $\pm$ 0.26\\
2452462.5905 & 60.645 & 159.26 & 16.80 $\pm$ 0.77 $\pm$ 0.26\\
2452462.5949 & 60.767 & 160.35 & 16.05 $\pm$ 0.68 $\pm$ 0.24\\
2452465.6268 & 61.541 & 170.27 & 16.76 $\pm$ 1.05 $\pm$ 0.26\\
2452470.6203 & 61.621 & 172.05 & 14.94 $\pm$ 0.44 $\pm$ 0.23\\
2452470.6234 & 61.650 & 172.82 & 15.59 $\pm$ 0.42 $\pm$ 0.24\\
2452470.6278 & 61.687 & 173.92 & 16.70 $\pm$ 0.44 $\pm$ 0.25\\
\hline
B3-M0\\
2452770.6605 & 133.838 & 59.85 & 1.379 $\pm$ 0.07 $\pm$ 0.02\\
2452770.6656 & 133.277 & 61.33 & 1.356 $\pm$ 0.14 $\pm$ 0.02\\
\hline
\end{tabular}
\begin{list}{}{}
\item[$^{\mathrm{*}}$] E0-G1 measurements
reported by Kervella et al.~(\cite{kervella03b}).
\end{list}
\end{table}

\section{Simulation of a 3D atmosphere\label{simu}}

In order to model the intensity profile of $\alpha$\,Cen~B, we have
performed  realistic, time dependent, three-dimensional radiative
hydrodynamical simulations  of its surface. The emerging intensity
of the atmospheric model in different directions is used to built
theoretical monochromatic limb darkening profiles for various
wavelengths covering the spectral domains of interest for the VINCI
and AMBER instruments (in the 1.0-2.4\,$\mu$m range).

\subsection{The stellar atmosphere modelling}

  The numerical code used for this work belongs to a new generation of 3D
  atmospheric codes developed for the study of solar { (e.g. Stein \& Nordlund \cite{stein89}, \cite{stein98})
  and stellar (e.g. Nordlund \& Dravins \cite{nordlund90}, Asplund et al. \cite{asplund99}, Allende-Prieto et al. \cite{allende02}, Ludwig et
al. \cite{ludwig02}) granulation and line formations (e.g. Asplund
et al. \cite{asplund00a}, \cite{asplund00b}, \cite{asplund00c},
\cite{asplund04}, \cite{asplund05}).}
 The code solves the non-linear, compressible equations of mass, momentum and energy conservation
 on an eulerian mesh. The 3D radiative transfer is solved at each time step along different inclined rays
 for  which we have assumed local thermodynamical equilibrium (LTE).
 We have considered 10 latitudinal $\mu$ points and 4 longitudinal $\varphi$ points, and
 checked that  a finer grid in $(\mu,\varphi)$ does not change the properties of the model.
 Realistic equation-of-state (including ionization, dissociation and recombination) and opacities
 (Uppsala opacity package, { Gustafsson et al. \cite{gus75})} are used. The line-blanketing
  is taken into account through the opacity binning technic (Nordlund ~\cite{nordlund82}).  In the present
   simulation we have considered
 a cartesian grid of  $(x,y,z)$ = $125\times125\times 82$ points. The geometrical sizes are $6\times 6$\,Mm
for the horizontal directions and 5\,Mm for the vertical one. The
dimensions of this domain are large enough to include a sufficiently
large number of granules (n $\geq$ 20) simultaneously { which
prevents} statistical bias. Periodic boundary condition is applied
for the horizontal directions and transmitting vertical boundaries
are used at the top and bottom of the domain. The base of the domain
is adjusted to have a nearly adiabatic, isentropic and featureless
convective transport. The upper boundary is placed sufficiently high
in the atmosphere so that it does not
influence the property of the model. \\
A detailed description of the current version of the code used in
this paper may be found in Stein \& Nordlund (\cite{stein98}).
Unlike 1D hydrostatic models that reduce all hydrodynamics into a
single adjustable parameter, the present simulations are done {\it
ab initio} by solving the complete set of RHD equations in a
self-consistent way. All the dynamics and turbulence of the model
come naturally from the equations of physics. Nothing is adjusted
like the convective flux in the MLT formalism. The diagnostic made
by such RHD simulations is therefore much more realistic than the 1D
models. We emphasize that the realism of these 3D simulations has
been intensively checked for solar line formations (e.g. Asplund et
al. \cite{asplund00b}, \cite{asplund00c}, \cite{asplund04}),
helioseismology (e.g. Rosenthal et al. \cite{rosen99}) and also for
stellar line formations (e.g. Allende-Prieto et al. 2002).

\begin{figure}
\includegraphics[ width=8cm]{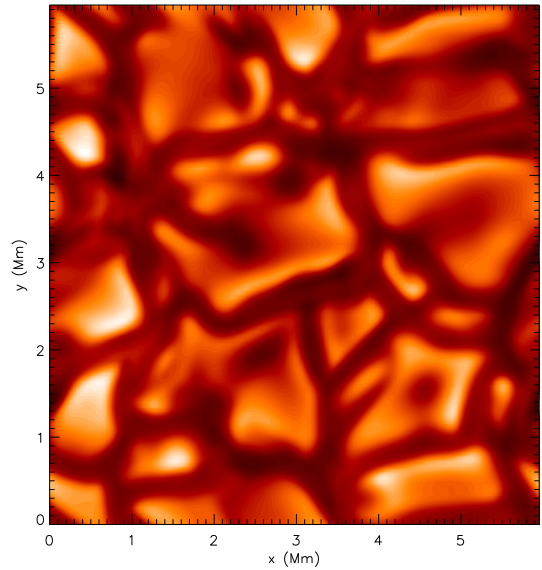}   
\caption{Snapshot of the disk-center ($\mu=1$) intensity emerging at
the stellar surface at a representative time.} \label{fig:granule}
\end{figure}

The adopted atmospheric parameters are those of Morel et al.
(\cite{morel00}), i.e. ${\rm T_{eff}} = 5260$ K, ${\rm log g} =
4.51$ and ${\rm [Fe/H]} = +\, 0.2$.
 The simulation has been run for a few
hours of stellar time which covers several convective turn-over
times. The result is a three-dimensional, time dependent box
representing the stellar surface. A snapshot of the disk-center
surface intensity is represented on Fig \ref{fig:granule}. The
structure of our model is similar to the one obtained
 by Nordlund \& Dravins (\cite{nordlund90}) but is even more  realistic since the present version
  of the code solves  compressible equations of hydrodynamics and uses more grid-points which
   allows a better treatment of the turbulence.

\subsection{3D limb darkening}

\begin{figure*}[ht!]
\centering \hbox{\includegraphics[ width=9cm]{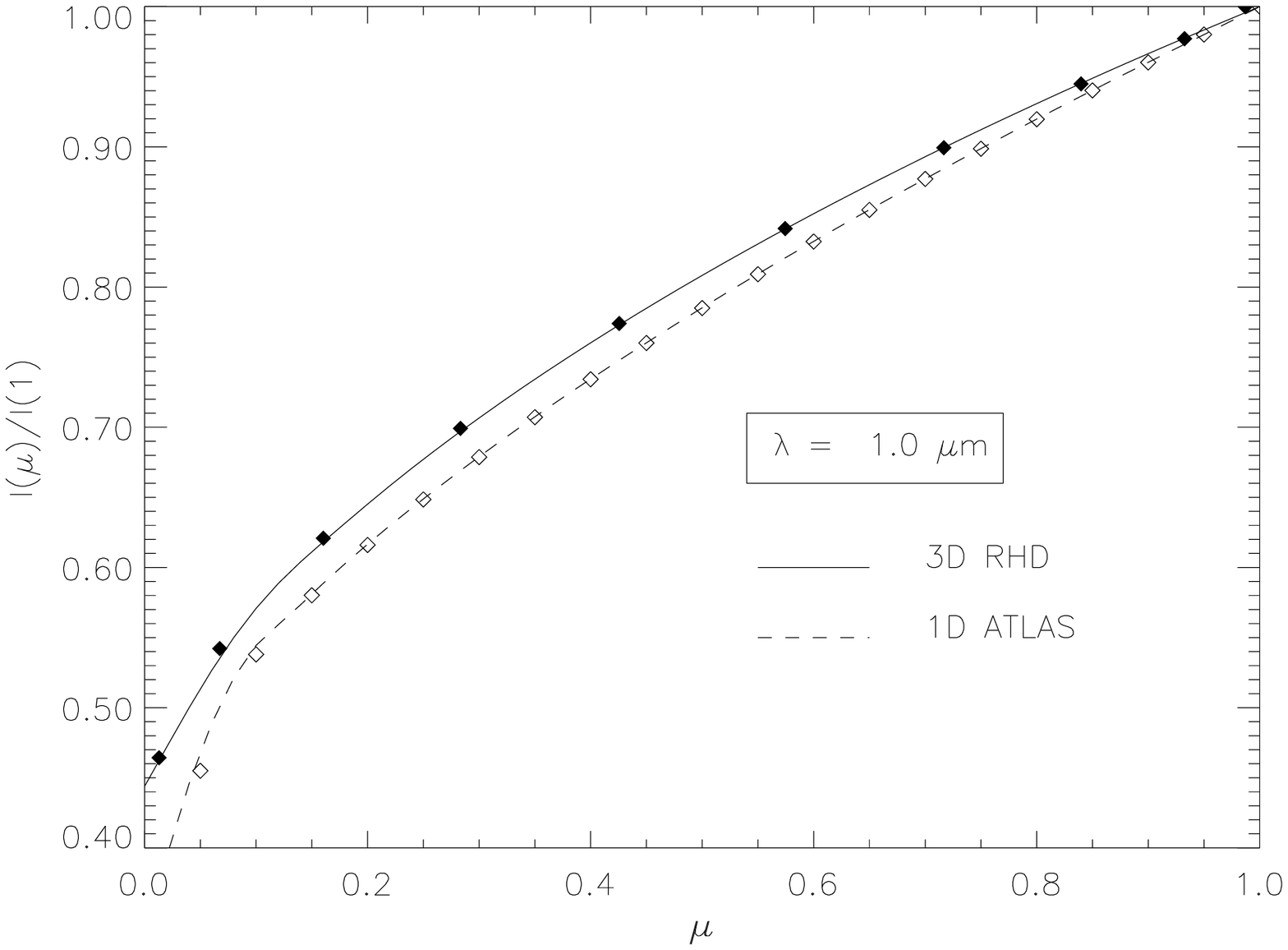}
\includegraphics[ width=9cm]{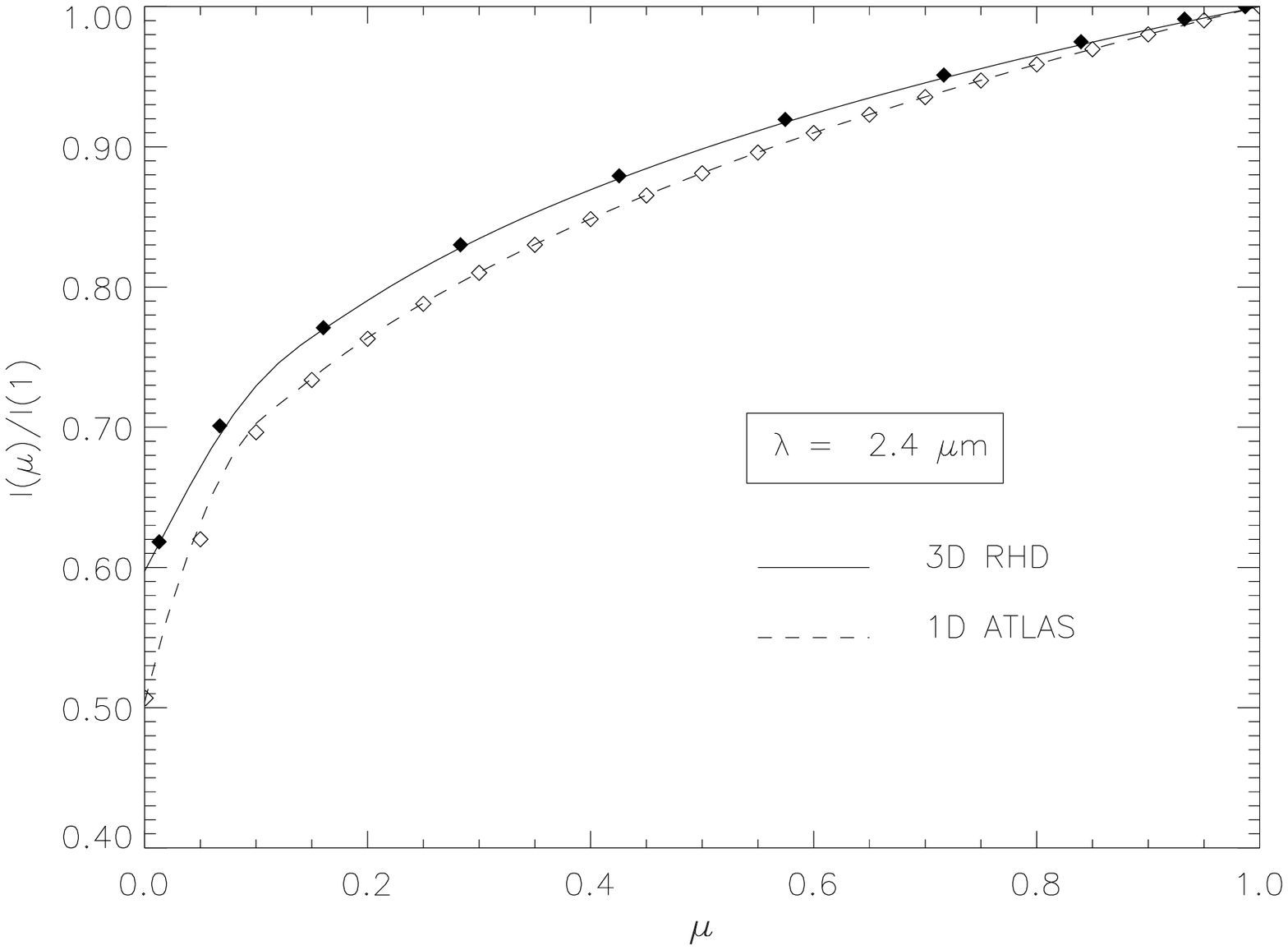} }
\caption{Normalized monochromatic center-to-limb variation
$I_\lambda(\mu)/I_\lambda(1)$ of the surface intensity obtained by
3D RHD simulations of $\alpha$ Cen B as a function of $\mu$ for two
different wavelengths: $1\,\mu$m (left panel) and $2.4\,\mu$m (right
panel), which correspond to the extreme limits of the spectral
domain we have considered in this paper. The solid lines represent
the 3D RHD limb darkening whereas the dashed lines represent limb
darkening derived from 1D ATLAS atmospheric models. In both case,
the symbols $\diamond$  represent the values extracted from both 1D
(white) and 3D (black) simulations.} \label{limbpred}
\end{figure*}

The monochromatic surface intensity is computed for various
latitudinal $\mu$ and longitudinal $\varphi$ directions at the
stellar surface. The limb darkening $I_\lambda (\mu)$ is obtained by
horizontal ($x,y$), longitudinal and time averages of the surface
intensity. For the time average we consider a sequence of 2 hours of
stellar time.
The results are plotted in Fig.~\ref{limbpred} for the two extreme
wavelengths of our spectral domain, 1.0 and 2.4\,$\mu$m. For
comparison, we overplot limb darkening obtained from a 1D ATLAS9
model for the same wavelengths  and for the same stellar fundamental
parameters. It appears that 3D RHD produces a { less} important
center-to-limb variation than a 1D static model. The departure from
1D model increases with decreasing wavelengths. Such behavior was
also found by Allende Prieto et al.~(\cite{allende02}) for Procyon.
However in the case of $\alpha$ Cen B, the departure from 1D to 3D
limb darkening is smaller, as a consequence of a less efficient
convection in K dwarfs as compared to F stars.

The reason why the emergent intensity differs between 1D and 3D
models is due to the fact that the properties of the superadiabatic
and surface convective layers cannot be well described by the mixing
length formalism, whatever parameter we choose. The temperature
inhomogeneities (granulation) together with the strong sensitivity
of the opacity ($H^{-}$) to the temperature make the warm ascending
flows more opaque than they would be for an homogeneous  1D model.
This purely { three-dimensional} effect added with  the contribution
of the turbulent pressure, pushes the location of the surface to
lower densities. The temperature gradient in these regions is
steeper than in the 1D case (see Nordlund \&
Dravins~\cite{nordlund90}). Since the continuum is formed in these
layers, the emergent intensity is different.

The correction due to 3D simulations (a few percents) is small but
not negligible with respect to the precision obtained by the new
generation of interferometric instruments like VINCI or AMBER. This
improvement is essential to derive an accurate angular diameter of
the star. We report in Table~\ref{limb_amb} our limb darkening
predictions for a series of continuum wavelengths between 1.0 and
2.4\,$\mu$m, corresponding to the $JHK$ range accessible to the
AMBER instrument.

\begin{table*}
 \caption{Limb darkening $I(\lambda, \mu)$  for various
wavelengths over the $JHK$ range.  }
 \label{limb_amb}
\begin{tabular}{c c c c c c c c c c c c }
\hline
      $\lambda$\,($\mu$m) / $\mu$    &        0.0  &   0.1  &   0.2  &   0.3 &    0.4  &   0.5 &    0.6
   &   0.7   &  0.8  &   0.9   &  1.0 \\
 \hline
1.050  & 0.4434   &  0.5745     & 0.6453   &  0.7069   & 0.7605 &
0.8087  &   0.8527  &   0.8932   &  0.9311   & 0.9667  &
1.0000 \\
 1.270   &  0.4646  &   0.6017    & 0.6738  &
0.7347  &   0.7860 &    0.8310  &   0.8711  &   0.9074   &
0.9406  &   0.9715  &   1.0000 \\
1.650    &  0.4838  & 0.6752 & 0.7487  &   0.8039  &   0.8462 &
0.8812  & 0.9110  & 0.9369   &  0.9601  &   0.9811  &   1.0000
\\
2.000 &  0.5442 &   0.7063  &   0.7707  &   0.8202 &    0.8585 &
0.8905 & 0.9178 &    0.9417  &   0.9630  &   0.9825 & 1.0000
\\
 2.200 & 0.5729   &  0.7220   &  0.7817  &   0.8283  & 0.8646
& 0.8950  &   0.9211  &   0.9439  &   0.9645  & 0.9831  &
1.0000 \\
2.400   &  0.5968    & 0.7353   &  0.7912   & 0.8352 & 0.8698 &
0.8988   &  0.9239   &  0.9458   & 0.9656  & 0.9836 & 1.0000 \\
 \hline
 \end{tabular}
\end{table*}

\section{Visibility model and angular diameter of $\alpha$\,Cen\,B}\label{discussion}

In this Section, we describe the application of our 3D limb
darkening models to the interpretation of the VINCI measurements of
$\alpha$\,Cen~B.

\subsection{ Limb darkened disk visibility model}\label{V_model}

In the simple case of a centro-symmetric star such as
$\alpha$\,Cen\,B, the visibility function measured using a broadband
interferometric instrument such as VINCI is defined by three
wavelength dependent functions:
\begin{enumerate}
\item The spectral energy distribution $S(\lambda)$ of the star, expressed
in terms of photons (VINCI uses a photon counting detector).
\item The wavelength dependent intensity profile of the star:
$I(\lambda,\mu)/I(\lambda,1)$.
\item The spectral transmission $T(\lambda)$ of the instrument, including
the atmosphere, all optical elements and the detector quantum
efficiency.
\end{enumerate}
Out of these three functions, $T(\lambda)$ is known from the
conception of the instrument, as well as calibrations obtained on
the sky (see Kervella et al.~\cite{kervella03b} for details). The
spectral energy distribution of the star $S(\lambda)$ can be
measured directly using a spectrograph, or taken from atmospheric
numerical models.

From the 3D RHD simulations presented in Sect.\,\ref{simu}, we have
obtained intensity profiles for ten distinct wavelengths over the K
band (chosen in the continuum). For each of these profiles, ten
values of $\mu$ were computed. The resulting $10 \times 10$ elements
2D table $I(\lambda,\mu)/I(\lambda,1)$ was then interpolated to a
larger $60\times 50$ elements table, with a 10\,nm step in
wavelength (over the 1.90-2.50\,$\mu$m range), and a 0.02 step in
$\mu$. This interpolation preserves well the smooth shape of the
intensity profile function.  This procedure was also used to build
the $I(\lambda,\mu)/I(\lambda,1)$ table based on 1D Kurucz model.
The original sample ($10\times 20$) was interpolated to the same
final grid as the 3D model.

We can derive the monochromatic visibility law $V_\lambda(B,
\theta)$ from the monochromatic intensity profile $I(\lambda,\mu)$
using the Hankel integral:
\begin{equation}
V_\lambda(B, \theta)  = \frac{1}{A} \int_0^1{I(\lambda,\mu)
J_0\left( \frac{\pi\,B\,\theta_{\rm LD}}{\lambda} \,\sqrt{1-\mu^2}
\right) \,\mu\,d\mu}\, ,
\end{equation}
where $B$ is the baseline (in meters), $\theta$ the limb darkened
angular diameter (in radians), $J_0$ the zeroth order of the Bessel
function, $\lambda$ the wavelength (in meters), $\mu = \cos \theta$
the cosine of the azimuth of a surface element of the star, and $A$
the normalization factor:
\begin{equation}
A = \int_0^1 I(\lambda, \mu)\,\mu\,d\mu\, ,
\end{equation}
To obtain the visibility function observed by VINCI in broadband, we
have to integrate this function taking into account the transmission
of the instrument and the spectral energy distribution of the star:
\begin{equation}
V_K(B,\theta)  = \frac{\int_K{ \left[V_\lambda(B, \theta)\
T(\lambda)\ S(\lambda)\right]^2\,\lambda^2\,d\lambda }} {\int_K{
\left[T(\lambda)\ S(\lambda)\right]^2\,\lambda^2 \,d\lambda }}\, ,
\end{equation}
Note the $\lambda^2$ term that is necessary as the actual
integration of the squared visibility by VINCI over the $K$ band is
done in the Fourier conjugate space of the optical path difference
(expressed in meters), and is therefore done in wavenumber $\sigma =
1/\lambda$. This corrective term ensures that the integration of the
spectrum of the star is done precisely in the same way as in the
instrument.

This formulation is very general, as it does not make any particular
assumption on the spectrum of the star, or on the wavelength
dependence of its intensity profile $I(\lambda,\mu)$.

\subsection{Fit of the interferometric data and angular diameter of $\alpha$\,Cen\,B}

Considering the model discussed in Sect.~\ref{V_model}, we now
derive the limb darkened angular diameter $\theta_{\rm LD}$ of
$\alpha$\,Cen\,B. It is obtained  by a standard $\chi^2$  {
analysis} of the data. We define the reduced $\chi^2$ of our fit by
\begin{equation}
\chi^{2}_{\rm red} = \frac{1}{N-n} \sum^{N}_{i=1}{\left(\frac{V_i^2
- V^2_{\rm model}}{\sigma_i}\right)^2}
\end{equation}
where $n$ is the number of variables ($n=1$ for our fit), $N$ the
total number of measurements, $i$ the index of a particular
measurement, and $\sigma_i$ the standard deviation of the
measurement with index $i$.

\begin{figure}
\includegraphics[ width=9cm]{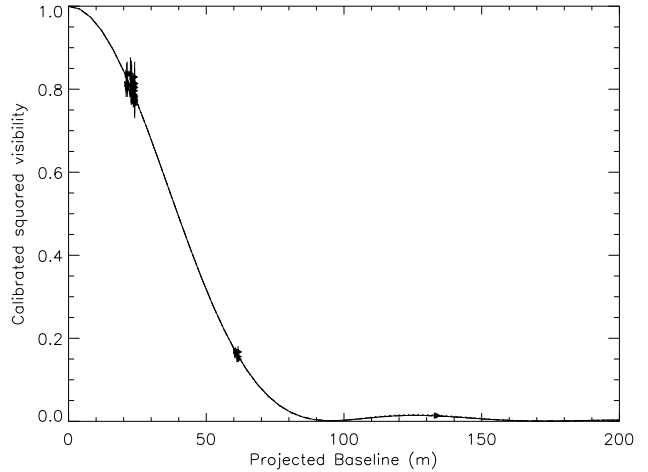}
\caption{Overview of $\alpha$\,Cen\,B squared visibilities. The
continuous line represents the broadband, limb darkened disk
visibility model derived from the 3D RHD, with $\theta_{\rm 3D} =
6.000 $\,mas.} \label{global_visib}
\end{figure}

\begin{figure*}
\centering \hbox{\includegraphics[ width=9cm]{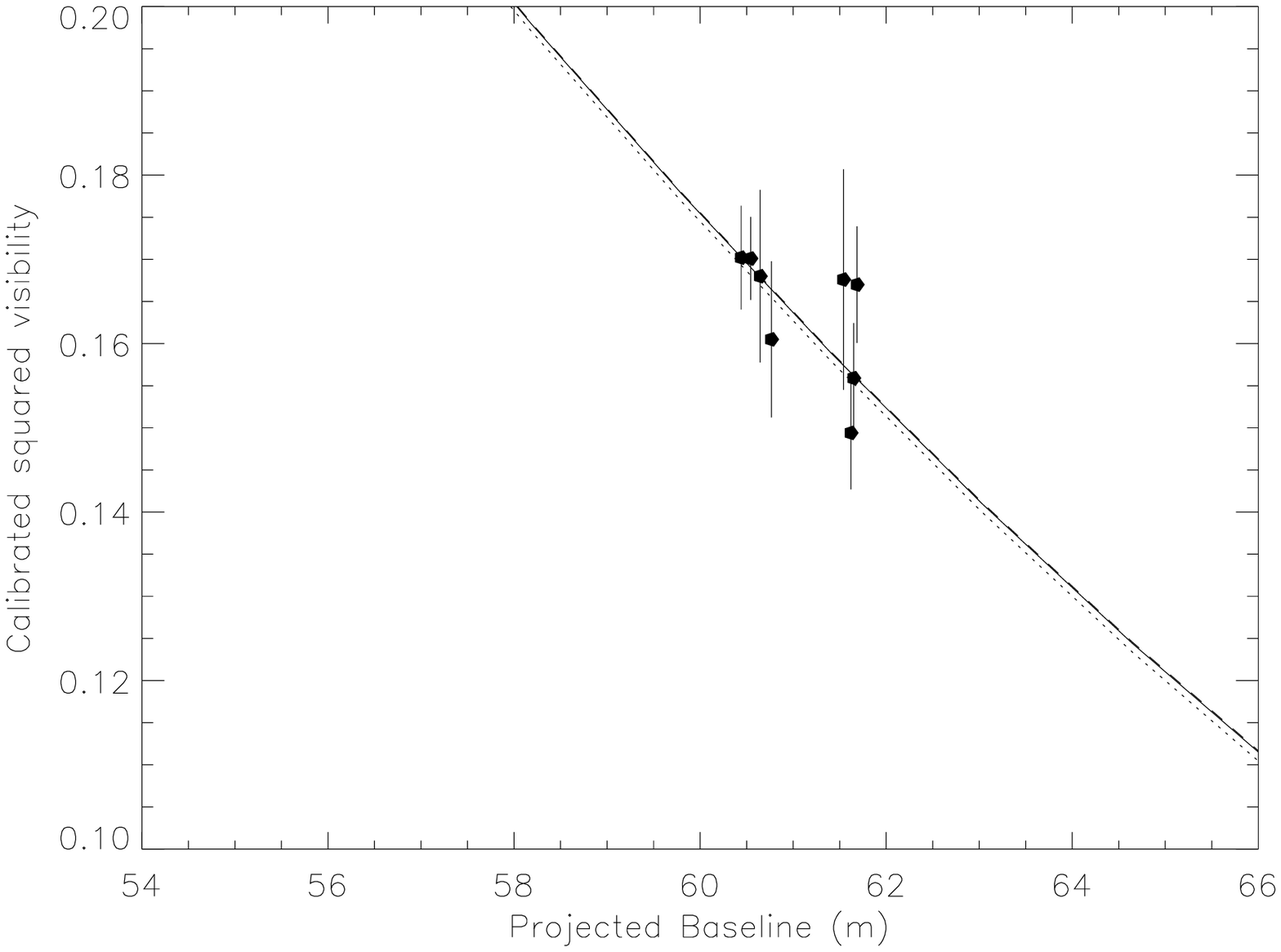}
\includegraphics[width=9cm]{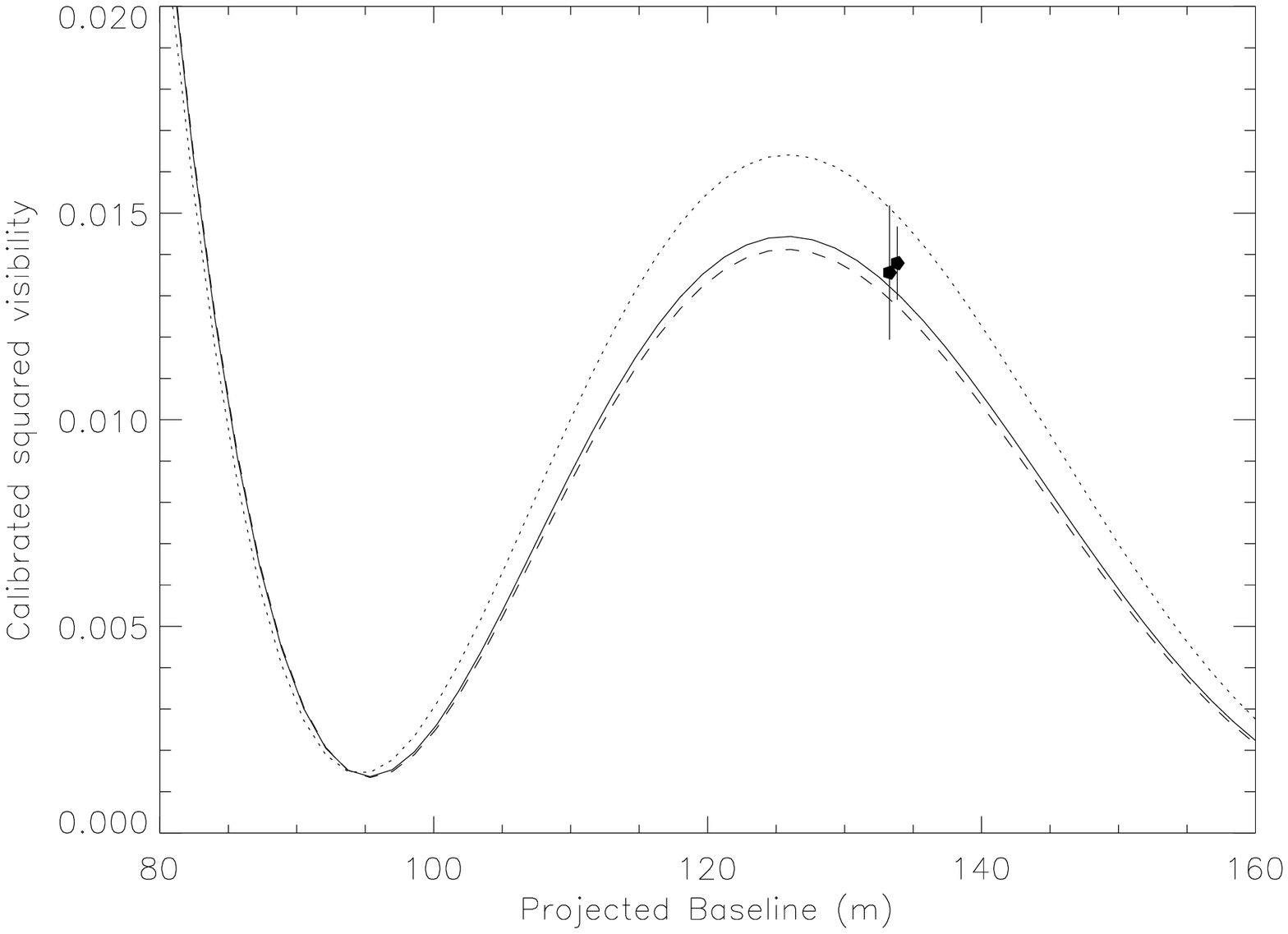} }\caption{Close up views of
the squared visibilities of $\alpha$\,Cen\,B in the lower part of
the first lobe (left panel) and the second lobe (right panel). The
continuous line represents the broadband, limb darkened disk
visibility model derived from the 3D RHD, with $\theta_{\rm 3D} =
6.000$ \,mas. The dashed lines correspond to results obtained from
1D ATLAS model with $\theta_{\rm 1D} = 6.017$\,mas. The upper dotted
curve is a UD model with $\theta_{\rm UD} = 5.881$ mas.  }
\label{detail_visib}
\end{figure*}

 The $\chi^2$ minimization was computed for three center-to-limb
models: uniform disk (UD), 1D ATLAS and 3D RHD. In each case the
broadband square visibility curve $V_K^2(B,\theta)$ is shown in
Fig.\,\ref{global_visib} and Fig.\,\ref{detail_visib}. In addition
to the purely statistical error, we must also take into account two
systematic error sources: the calibration uncertainty and the
wavelength uncertainty. The calibration uncertainty comes from the
errors on the {\it a priori} angular sizes of the five calibrators
that were used for the VINCI observations. It amounts to 0.012\,mas
on the final angular diameter. The wavelength uncertainty comes from
the imperfect knowledge of the transmission of VINCI, in particular
of its fluoride glass optical fibers. This transmission was
calibrated on sky (Paper~I), and the uncertainty on this measurement
is estimated to be 0.15\%. As it impacts linearly on the angular
diameter value, it corresponds to 0.009\,mas. These two systematic
factors add up quadratically to the 0.013\,mas statistical
uncertainty, and result in a total error of 0.021\,mas on the
angular diameters of $\alpha$\,Cen\,B.
The best fit angular diameter that we derive using our 3D limb
darkening model is $\theta_{3D} = 6.000 \pm 0.021$\,mas. The 1D
model produces a slightly larger diameter, $\theta_{1D} = 6.017 \pm
0.021$\,mas and the UD disk produces naturally a much smaller
diameter, $\theta_{\rm UD} = 5.881 \pm 0.021$\,mas.  \\
There is no significant departure between the three models in the
first lobe of visibility. However, different amplitudes of the
second lobe are observed. While UD model produces higher
visibilities, the 1D limb darkened model leads to slightly too low
visibilities compared to our observations. Overall, the  3D model
leads to a slightly better agreement with observations.

As expected, the difference 3D/1D is rather small since we are
working in the near-infrared (K-band) and for a dwarf star. It is
nonetheless comparable with  $\sigma_{stat}$, and therefore
significant.

\subsection{Linear diameter}

Assuming the parallax value of S\"oderhjelm~(\cite{soderhjelm99}),
$\pi = 747.1 \pm 1.2$\,mas\footnote{One should note that there is a
rather broad distribution of parallax values for $\alpha$\,Cen in
the literature, as discussed in Paper I. While the value from
S\"oderhjelm~(\cite{soderhjelm99}) is the result of a careful
reprocessing of the {\it Hipparcos} data, the possibility of a bias
beyond the stated 1\,$\sigma$ error cannot be completely excluded,
in particular due to the extreme brightness and binarity of
$\alpha$\,Cen.}, we find a linear radius of $0.863 \pm
0.003$\,R$_{\odot}$ which agrees with results presented in Paper~1.
We estimate that the adopted uncertainty  in the {\it a priori}
assumed (of ${\rm T_{eff}} = 50$ K) leads to an error of about
$0.0003$ $R_{\odot}$, i.e. ten times smaller than the derived
uncertainties. From the 1D analysis, we derive a radius of $0.865
\pm 0.003$\,R$_{\odot}$, larger than the radius found by the RHD
approach by about $1\,\sigma_{stat}$. In addition to the corrections
it provides, the use of 3D simulations is also motivated by the
absence of adjustable parameters, which is not the case for 1D
models.

This slightly smaller linear radius obtained from 3D RHD simulations
compared with the one derived from 1D ATLAS model supports the
suggestion of a  smaller mass ($M=0.907\,M_{\odot}$, Kervella et al.
2003) rather than the larger one ($M=0.934 \pm 0.007 \,M_{\odot}$)
proposed by Pourbaix et al.\,(\cite{pnm02}). However, stellar
evolution models are sensitive to many parameters and a smaller
radius do not always lead to a smaller mass.  More investigations
are thus needed before we can reach a definite conclusion about the
mass of $\alpha$ Cen B. In this context, our improved radius
provides an additional constraint on asteroseismic diagnostics.

\section{Conclusion}

In this paper we improve the determination of the radius of $\alpha$
Cen B in two respects. Firstly, we report the first interferometric
measurements in the second lobe of visibility. Secondly, in order to
derive a reliable value of the angular diameter of the star, we have
performed realistic 3D RHD simulations of the surface of $\alpha$
Cen B. By comparison with observations we find a radius of $0.863
\pm 0.003$ $R_{\odot}$. The correction provided by the 3D approach
is { small but} significant (especially in the $K$ band probed with
VINCI) since it provides a radius smaller by roughly $1
\,\sigma_{stat}$  compared with what can be obtained by 1D
 models.
 Moreover, the use
of 3D RHD is preferable since it does not introduce { adjustable}
 parameters { to describe convection}. We also emphasize
that for hotter A-F stars the correction due to 3D analysis will be
larger than for $\alpha$\,Cen~B. Though it is small, we have shown
that even for a K-dwarf like $\alpha$ Cen B, the correction obtained
by the use of RHD simulations should not be neglected for very high
precision interferometric measurements. In the next few years, the
combination of high visibility precision and long baselines will
require the use of realistic theoretical models of the stellar limb
darkening to extract the true photospheric angular diameter of the
observed stars from the observed visibilities. Conversely,
observations beyond the first minimum of the visibility function
will sample directly the light distribution on the surface of the
stars, therefore providing constraints for the atmosphere structure
models. Future observations with the VLTI will allow to sample
tightly the second lobe of the visibility function of several solar
analogs (including $\alpha$\,Cen~A and B), and therefore derive
their intensity profiles with high accuracy. The comparisons between
our theoretical predictions of limb darkening and the future
observations made by AMBER will be an excellent test for our
modelling of the surface of $\alpha$\,Cen~B. Indeed, AMBER will
provide new interferometric data which will contain much more
information compared with VINCI. There will be two major advantages
with AMBER:
 \begin{itemize}
\item It will provide a wavelength dependence of the visibility ([1.9-2.4]
$\mu$m) therefore allowing differential studies of the limb
darkening as a function of wavelength.
\item AMBER can combine simultaneously the light from three telescopes and
therefore measures the closure phase. This gives an advantage to
determine the angular size of the star when observing in the minima
of the visibility function.
\end{itemize}
These improvements will lead to better constrained angular diameters
of $\alpha$\,Cen~A and B, and therefore a high precision measurement
of the ratio of the linear radii of A and B, independent of the
parallax.

\begin{acknowledgements}
We thank Vincent Coud\'e du Foresto for important remarks that led
to improvements of this paper for an early stage of this paper. We
thank the anonymous referee for constructive remarks. These
interferometric measurements have been obtained using the VLTI (ESO
Paranal, Chile), and were retrieved from the ESO/ST-ECF Archive
(Garching, Germany). LB thanks CNES for financial support and {\AA}.
Nordlund for providing his RHD code. We also thank Claude Van't Veer
for providing the ATLAS model.
\end{acknowledgements}


\begin{thebibliography}{}
\bibitem[2002]{allende02} Allende Prieto, C., Asplund, M., Garcia L\`opez, R. J. \& Lambert, D. L. 2002, ApJ, 567, 544
\bibitem[2004]{allende04} Allende Prieto, C., Asplund, M., Fabiani Bendicho, P. 2004, A\&A, 423, 1109
\bibitem[2000]{asplund99} Asplund, M., Nordlund, {\AA}, Trampedach, R. Stein, R. F. 1999, A\&A, 346L, 17
\bibitem[2000a]{asplund00a} Asplund, M., Ludwig, H.-G., Nordlund, {\AA}., \& Stein, R. F. 2000a, A\&A, 359, 669
\bibitem[2000b]{asplund00b} Asplund, M., Nordlund, {\AA}., Trampedach, R., Allende Prieto, C., Stein, R. F. 2000b, A\&A, 359, 729
\bibitem[2000c]{asplund00c} Asplund, M., Nordlund, {\AA}.,  Trampedach, R., Stein, R. F. 2000c, A\&A, 359, 743
\bibitem[2004]{asplund04} Asplund, M., Grevesse, N., Sauval, A. J., Allende Prieto, C., Kiselman, D. 2004, A\&A, 417, 751
\bibitem[2005]{asplund05} Asplund, M., Grevesse, N., Sauval, A. J., Allende Prieto, C., Blomme, R. 2005, A\&A, 431, 693
\bibitem[2002]{borde02} Bord\'e, P., Coud\'e du Foresto, V., Chagnon, G. \& Perrin, G. 2002, A\&A, 393, 183
\bibitem[2001]{bouchy01} Bouchy, F., Carrier, F. 2001, A\&A, 374, L5
\bibitem[2001]{bouchy02} Bouchy, F., Carrier, F. 2002, A\&A, 390, 205
\bibitem[2003]{carrier03} Carrier, F., Bourban, G. 2003, A\&A, 406, 23
\bibitem[2000]{claret00} Claret, A. 2000, A\&A, 363, 1081
\bibitem[1999]{cohen99} Cohen, M., Walker, R. G., et al. 1999, AJ, 117, 1864
\bibitem[2004]{egge04} Eggenberger, P., Charbonnel, C., Talon, S., et al. 2004,  A\&A, 417, 235
\bibitem[1975]{gus75} Gustafsson, B., Bell, R. A., Eriksson, K., Nordlund, {\AA}.  1975, A\&A, 42, 407
\bibitem[1999]{hauschildt99} Hauschildt, P. H., Allard, F., Baron, E. 1999, ApJ, 512, 377
\bibitem[1998]{hestroffer98} Hestroffer, D., Magnant, C. 1998, A\&A, 333, 338
\bibitem[2000]{kervella00} Kervella, P., Coud\'e du Foresto, V., Glindemann, A., Hofmann, R. 2000, SPIE, 4006, 31
\bibitem[2003a]{kervella03a} Kervella, P., Gitton, Ph., S\'egransan, D., et al. 2003a, SPIE, 4838, 858
\bibitem[2003b]{kervella03b} Kervella, P., Th\'evenin, F., S\'egransan, D., et al. 2003b (Paper I), A\&A, 404, 1087
\bibitem[2004]{kervella04} Kervella, P., S\'egransan, D. \& Coud\'e du Foresto, V. 2004, A\&A, 425, 1161
\bibitem[1992]{kurucz92} Kurucz, R. L. 1992, IAU Symp.~149: The Stellar Populations of Galaxies, 149, 225
\bibitem[2002]{ludwig02} Ludwig, H., Allard, F., Hauschildt, P.H. 2002,  A\&A, 395, 99
\bibitem[2000]{morel00} Morel, P., Provost, J., Lebreton, Y., Th\'evenin, F.,  Berthomieu, G. 2000, A\&A, 363, 675
\bibitem[1994]{neckel94} Neckel, H., Labs, D. 1994, SoPh, 153, 91
\bibitem[1982]{nordlund82} Nordlund, {\AA}., A\&A 1982, 107, 1
\bibitem[1990]{nordlund90} Nordlund, {\AA}., Dravins, D. 1990, A\&A, 228, 155
\bibitem[2000]{petrov00} Petrov, R., Malbet, F., et al. 2000, SPIE, 4006, 68
\bibitem[1977]{pierce77} Pierce, A. K., Slaughter, C.D. 1977, SoPh, 51, 25
\bibitem[2002]{pnm02} Pourbaix, D., Nidever, D., McCarthy, C., et al. 2002, A\&A, 386, 280
\bibitem[2004]{robbe04} Robbe-Dubois, S., Petrov, R. G., Lagarde, S., et al. 2004, SPIE, 5491, 1089
\bibitem[1999]{rosen99} Rosenthal, C. S., Christensen-Dalsgaard, J., Nordlund, {\AA}., Stein, R.
F.,  Trampedach, R. 1999, A\&A, 351, 689
\bibitem[1989]{stein89} Stein R.F., Nordlund {\AA}. 1989, ApJL, 342, 95
\bibitem[1998]{stein98} Stein R.F., Nordlund {\AA}. 1998, ApJ, 499, 914
\bibitem[1999]{soderhjelm99} S\"oderhjelm, S. 1999, A\&A, 341, 121
\bibitem[2002]{the02} Th\'evenin, F., Provost, J., Morel, P., et al. 2002,  A\&A, 392, L9
\bibitem[2004]{thoul04} Thoul, A., Scuflaire, R., Noels, A., et al. 2003,  A\&A, 402, 293
\end{thebibliography}
\end{document}